\documentclass[aps,prd,twocolumn,superscriptaddress,bibnotes,longbibliography,preprintnumbers,floatfix,10pt]{revtex4-2}

\usepackage{graphicx}
\usepackage{xcolor}
\usepackage{amsmath,amsfonts,amsthm,amssymb}
\usepackage[colorlinks]{hyperref}
\usepackage{mathtools}
\usepackage{bm}
\usepackage{multirow}
\usepackage[normalem]{ulem}
\usepackage{float}
\usepackage[noabbrev,capitalise]{cleveref}
\makeatletter
\AddToHook{cmd/appendix/before}{\def\cref@section@alias{appendix}}
\makeatother

\graphicspath{{./figures/}}

\newcommand{\orcid}[1]{\href{https://orcid.org/#1}{\includegraphics[width=10pt]{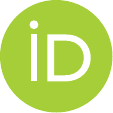}}}

\newcommand{\ror}[1]{\href{https://ror.org/#1}{\includegraphics[width=10pt]{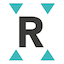}}}

\hypersetup{
    pdfnewwindow=true,      
    colorlinks=true,       
    linkcolor=violet,          
    citecolor=violet,        
    filecolor=violet,      
    urlcolor=violet        
}

\begin{document}
\preprint{CETUP 2025-005}
\title{New Multi-messenger Probe of Dark Matter-Nucleon Interactions from \\
Ultra-high Energy Cosmic Ray Acceleration}

\author{Stephan A. Meighen-Berger \orcid{0000-0001-6579-2000}\,}
\email{stephan-meighen-berger@uiowa.edu}
\affiliation{School of Physics, The University of Melbourne, Victoria 3010, Australia \ror{01ej9dk98}}
\affiliation{Center for Cosmology and AstroParticle Physics (CCAPP), Ohio State University, 
Columbus, OH 43210, USA \ror{00rs6vg23}}
\affiliation{Department of Physics and Astronomy, University of Iowa, Iowa City, IA 52242, USA \ror{036jqmy94}}

\author{P. S. Bhupal Dev \orcid{0000-0003-4655-2866}}
\email{bdev@wustl.edu}
\affiliation{Department of Physics and McDonnell Center for the Space Sciences, Washington University, St. Louis, MO 63130, USA \ror{01yc7t268}}

\author{Matheus Hostert \orcid{0000-0002-9584-8877}}
\email{matheus-hostert@uiowa.edu}
\affiliation{Department of Physics and Astronomy, University of Iowa, Iowa City, IA 52242, USA \ror{036jqmy94}}


\begin{abstract}

It has been suggested that the density of dark matter (DM) halo can be highly enhanced around supermassive black holes at the centers of massive galaxies.
If real, these DM \emph{spikes} would offer new opportunities to probe the properties of DM. 
In this work, we point out that DM spikes can significantly impact the composition and survivability of ultra-high-energy cosmic rays accelerated near supermassive black holes. 
A large DM-nucleon cross section would fragment heavy nuclei into lighter elements and prevent them from attaining the energies observed at Earth. 
While the origin of cosmic rays remains a mystery, we show that if the highest-energy cosmic rays on Earth come from sources like NGC 1068, then cross sections of size $\sigma_{\chi p} \leq  10^{-33} \left( \frac{m_\chi}{\mathrm{GeV}}\right)\;\mathrm{cm^{2}}$ would be excluded by cosmic ray data. 
These bounds can be competitive with other existing probes and rule out new parameter space in the DM mass region $m_\chi\in [3\;\mathrm{MeV}, 30\;\mathrm{MeV}]$. 
While the uncertainties on the acceleration mechanism of cosmic rays prevent us from setting robust limits, our study highlights an important connection between DM spikes and cosmic ray physics that is complementary to existing cosmological and direct detection constraints. 

\end{abstract}

\maketitle


\section{Introduction}\label{sec:intro}

Supermassive black holes (SMBH), believed to exist at the centers of almost all massive galaxies,  provide extreme and naturally-occurring laboratories with which to probe the microphysics of cosmic rays (CRs), neutrinos, and dark matter (DM).
These objects, with masses ranging from millions to billions of solar masses, create some of the most energetic environments in the Universe, where particles can be accelerated to ultra-high energies through various mechanisms including magnetic reconnection, shock acceleration, and interactions with accretion disk turbulence~\cite{Ptitsyna:2015nta,  Inoue:2019yfs, Murase:2019vdl, Katsoulakos:2020rkr, Kheirandish:2021wkm, Eichmann:2022lxh, Inoue:2022yak, Fiorillo:2023dts, Fiorillo:2024akm, Karavola:2024uui}.
The recent discovery of NGC 1068~\cite{IceCube:2022der} as a high-energy neutrino source and its lack of corresponding TeV gamma-ray signal have shown that the close vicinity of the central SMBH is the most likely source of high-energy neutrino production~\cite{Padovani:2024ibi}. 
The suppression of gamma rays, combined with the fact that this is the first steady source of neutrino observed, suggests that the emission region is deeply embedded within dense material that effectively absorbs high-energy electromagnetic radiation while allowing neutrinos to escape.
This, in turn, implies that CRs are accelerated near the central SMBH, within a few to hundreds of Schwarzschild radii from the event horizon.
These small distances would imply that the CR acceleration is taking place in an environment with a potentially large ambient DM density, assuming a cuspy profile. 

\begin{figure}[t!]
\centering
\includegraphics[width=\columnwidth]{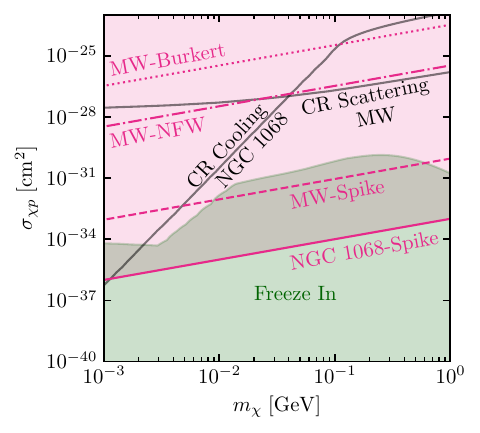}
\caption{The approximate bounds set here (pink region) using a spiked DM-density profile~\cite{Gondolo:1999ef} in an NGC 1068-type galaxy compared to other proposed bounds using CRs (black lines). Specifically, we are showing scattering~\cite{Cappiello:2018hsu}, and cooling~\cite{Herrera:2023nww} bounds. 
We also show DM-proton cross section values that can produce the correct relic abundance via freeze-in (green region)~\cite {Elor:2021swj, Bhattiprolu:2022sdd}. 
Our potential bounds using a Burkert~\cite{Burkert:1995yz}, NFW~\cite{Navarro:1995iw}, and spiked-density profiles for Milky-Way (MW)-like galaxies are shown as dotted, dashed-dotted, and dashed lines, respectively.}
\label{fig:sketch}
\end{figure}

Gamma-ray observations from other galaxies (e.g., NGC 4486) have also demonstrated that high-energy emission can originate from regions very close to the event horizon, supporting the plausibility of particle acceleration at these scales~\cite{Berge:2006uls, Aharonian:2007ig, Acciari:2008ah, HESS:2011huh, MAGIC:2012gcb, Aleksic:2014xsg}.
Such obscured, Compton-thick sources have long been theorized~\cite{Berezinsky:1977, Eichler:1979yy, Silberberg:1979hd, Hickox:2018xjf}, but their existence is only now being confirmed with the advent of multi-messenger astronomy~\cite{ IceCube:2021pgw, Giommi:2021bar, Fang:2022trf}. 
For instance, these obscured Active Galactic Nuclei (AGNs) have been used to explain the X-ray background (XRB) measurements~\cite{setti1989active}, specifically the peak between 20-30 keV~\cite{Comastri:1994bz, Treister:2006ec, Akylas:2012am, Ananna:2018uec}. 
Interestingly, non-jetted AGNs have also been used to explain the XRB~\cite{Gilli:2006zi} and astrophysical neutrino background~\cite{Padovani:2024tgx}. 
Similar arguments have also been recently applied to NGC 7469~\cite{Yang:2025lmb}.

While cuspy profiles are often used to describe DM halos, the presence of an SMBH can further enhance the DM density in its vicinity.
It has been suggested that the DM density profile could form a `spike' as we go closer to the SMBH due to adiabatic growth of the black hole~\cite{Gondolo:1999ef}. In other words, as the central black hole forms and grows, its gravity pulls in ambient DM, compressing it into a much steeper density profile, referred to as a ``spike". 
The origin of a spiked profile can be theoretically motivated simply from angular momentum conservation. With more mass accreting onto the SMBH, the minimum of the effective gravitational potential shifts towards the center, thus giving rise to a spiked profile. 
 
Qualitatively, the spike formation can be understood as follows: Starting from an initial DM density profile $\rho(r)\propto r^{-\gamma}$ with $\gamma$ an initial density slope (typically $0.5 \lesssim \gamma \lesssim 1.5$~\cite{Gnedin:2003rj, 10.1093/mnras/stv058, 2017PDU....15...90I, Hooper:2016ggc, Baumgart:2025dov}), the adiabatic growth of the black hole causes the orbits of DM particles to contract towards the center.
This process, which conserves adiabatic invariants, results in a new power-law density profile within the spike, $\rho^{\gamma_{\rm sp}}_{\rm sp}(r)\propto r^{-\gamma_{\rm sp}}$. 
The relationship between the initial slope $\gamma$ and the spiked slope $\gamma_{\rm sp}$ is given by $\gamma_{\rm sp}=\frac{9-\gamma}{4-\gamma}$~\cite{Gondolo:1999ef, Quinlan:1994ed}. For typical halo models where $\gamma$ ranges from 0 to 2, this transformation yields very steep inner profiles, with $\gamma_{\rm sp}$ values between 2.25 and 2.5. 
The spike is theorized to extend from a minimum radius, influenced by annihilation or capture effects, out to a characteristic radius $R_{\rm sp}\sim 1\;\mathrm{pc}$. 
These results were then improved by using a fully relativistic treatment~\cite{Sadeghian:2013laa}. 
The potential existence of these spikes has been leveraged in numerous analyses to enhance expected sensitivities for DM searches, including studies by FGST and PAMELA \cite{Sandick:2010yd}, various gamma-ray observations \cite{Regis:2008ij, Gorchtein:2010xa, Balaji:2023hmy}, and stellar observations~\cite{Lacroix:2018zmg, Gustafson:2025ypo, Acevedo:2025rqu}. Furthermore, the spike has been utilized to set bounds on DM interactions through its capture and annihilation within stars \cite{John:2023knt} and to predict the existence of dark stars \cite{BetancourtKamenetskaia:2025hgr}. Refs.~\cite{Wang:2021jic, Granelli:2022ysi} have used spikes around the TXS 0506+056 to constrain blazar-boosted DM. More recently, Ref.~\cite{Akita:2025dhg} used the spike to explain the neutrino signal from NGC 1068 via DM annihilation. Additionally, Refs.~\cite{Cline:2022qld, Ferrer:2022kei,Cline:2023tkp, DeMarchi:2024riu, Zapata:2025huq, DeMarchi:2025xag,Dev:2025tdv,  DeMarchi:2025uoo} have used spiked profiles to constraint neutrino-DM interactions. Collectively, these studies highlight the broad phenomenological relevance of spiked profiles. 
Note, however, that the existence of the spike is not yet established observationally and is still under debate~\cite{Ullio:2001fb, Lacroix:2018zmg, Shen:2023kkm}. See \Cref{app:arguments} for a review of possible mechanisms that could weaken or destroy such spikes.

In this paper, we discuss how the existence of a spiked DM halo impacts CR acceleration to ultra-high energies.
These ultra-high energy CRs (UHECRs) are expected predominantly to consist of heavier elements, including iron~\cite{Gaisser:2011klf, Thoudam:2016syr,Muzio:2019leu, Ivanov:2020rqn, PierreAuger:2023bfx, Ehlert:2023btz,Zhang:2024sjp, TelescopeArray:2024oux, IceCubeCollaborationSS:2025jbi, PierreAuger:2025eun, IceCube:2025baz}. 
A single scattering event between such energetic CRs and a DM particle during the acceleration process would cause the heavy  (iron) nucleus to shatter. 
If such fragmentation occurs, CRs would be unable to reach the observed ultra-high energies, particularly near the Greisen-Zatsepin-Kuzmin (GZK) cutoff~\cite{Greisen:1966jv, Zatsepin:1966jv}. In fact, the observation of several super-GZK events~\cite{HIRES:1994ijd,PierreAuger:2022qcg,TelescopeArray:2023sbd}  strongly suggests that (i) these UHECRs are most likely heavy nuclei (with higher GZK cutoff) rather than protons, and (ii)  fragmentation of these heavy nuclei is unlikely to happen at the source.

Here we ask the question: \textit{Given a DM-proton interaction, can high-energy CRs be accelerated within the DM spike to the ultra-high energies necessary to explain current observations by CR experiments \cite{Ivanov:2020rqn, PierreAuger:2020qqz}?} We split this question into three parts:

\begin{enumerate}
    \item Can CRs be accelerated to the required energies within an AGN?
    \item Can CRs be accelerated within their trapped region?
    \item Can the CRs escape their host galaxy?

\end{enumerate}
Given the observation of UHECRs on Earth, we can set bounds on the DM-nucleon cross section by requiring that the interaction length of CR-DM interactions is at least larger than the acceleration region.
\cref{fig:sketch} shows the approximate results we predict compared to other CR bounds. 
Not shown are bounds due to changes in the expected metallicity~\cite{Lu:2023aar} measured by CR experiments such as AMS~\cite{AMS:2021lxc}. There, it is assumed the DM-nucleon cross section behaves as the proton-nucleon cross section, scaled by a phenomenological factor $b_\chi$, $\sigma_{\chi  N} = b_\chi \sigma_{p  N}$. For instance, in some scenarios, the DM-nucleon cross section scales as $\sigma_{\chi N} \sim A^{2/3} \ln^2(s/{\rm GeV}^2)$, where $A$ is the nucleon count, and $\sqrt{s}$ the center-of-mass energy~\cite{Froissart:1961ux, Martin:1962rt, Glauber:1970jm, Block:2005pt, Ulrich:2010rg, Gaisser:2013bla}. With such a scaling, our bounds would \textit{improve} by $\mathcal{O}(100)$ when compared with Ref.~\cite{Lu:2023aar} and by $\mathcal{O}(1000)$ when compared with direct detection experiments. Since we wish to remain agnostic to the underlying model, we assume $\sigma_{\chi  N}$ is constant with the momentum transfer.

Note that the method presented here is similar to Refs.~\cite{Cappiello:2018hsu, Lu:2023aar}. The main difference is that those  authors focused on changes in the energy spectrum (at lower energies) due to multiple scattering. In contrast, here we focus on single-scatter events at higher energies.

This article is structured as follows.
In \cref{sec:dm_densities}, we briefly discuss the DM-density profiles used in this work. In \cref{sec:dm_cr_interactions} we discuss general CR-DM interactions. 
Then, in \cref{sec:bounds}, we derive bounds on DM-proton interaction rate using EeV CR iron nuclei by requiring that they do not fragment during acceleration to these energies.
We present our conclusions in \cref{sec:conclusion}.


\section{DM Densities}
\label{sec:dm_densities}

\begin{figure}[t]
\centering
\includegraphics[width=\columnwidth]{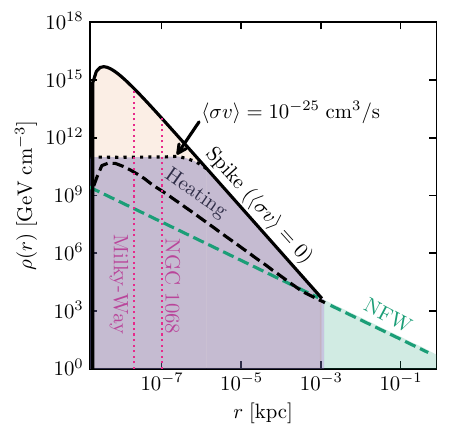}
\caption{The DM density as predicted by Ref.~\cite{Gondolo:1999ef} combined with the NFW~\cite{Navarro:1995iw} profile starting at $r = 4R_\mathrm{S} = 1.6\times 10^{-9}\;\mathrm{kpc}$ for an SMBH mass of $4\times 10^6M_\odot$. 
The orange region shows where DM annihilation could reduce the expected density. 
The black dashed line denotes the expected DM density when accounting for significant stellar heating~\cite{Balaji:2023hmy}, and the dotted black line indicates the plateau induced by DM  annihilation of $\langle\sigma v \rangle \sim 10^{-25}$~cm$^3$/s. While the density profile shown is for an MW-like galaxy, the pink vertical lines indicate the conservative acceleration regions of CRs for an MW-like or NGC 1068-like galaxy, respectively, based on multi-messenger observations.
}
\label{fig:density}
\end{figure}

To model the DM density profile near the SMBH, we follow the formalism of Ref.~\cite{Gondolo:1999ef}. We adopt a generalized NFW profile~\cite{Navarro:1995iw} for the ambient DM halo outside of the spike region, with an initial slope parameter $\gamma$ set to 1 and $\rho_0 = 0.4\;\mathrm{GeV / cm^3}$, the local ($r_0 = 8.5\;\mathrm{kpc}$) DM density. In the simplest scenario, where the central SMBH grows adiabatically, via the slow in-fall of gas, the DM density can be written as~\cite{Gondolo:1999ef}
\begin{equation}\label{eq:adiabatic}
    \rho_{\rm{sp}}^{\gamma_{\rm{sp}}} (r) \approx \rho_0 \left(\frac{R_{\rm{sp}}}{r_0}\right)^{-\gamma} \left(1 - 4\frac{R_{\rm S}}{r}\right)^3 \left(\frac{R_{\rm{sp}}}{r}\right)^{-\gamma_{\rm{sp}}}.
\end{equation}
Here $R_{\rm S}=2 G M_{\rm BH}$ is the Schwarzschild radius of the black hole, $R_{\rm sp} =\alpha_\gamma r_0 (M_{\rm{BH}} / (\rho_0 r^3))^{1 / (3 - \gamma)}$ the size of the spike, and $\alpha_\gamma$ the normalization to match the density outside of the spike. Here $\gamma = 1$ corresponds to $ \alpha_\gamma = 0.12, \gamma_{\rm{sp}} = 7/3$. Thus, the full density profile is
\begin{equation}
    \rho_\chi (r) = 
    \begin{cases}
        \displaystyle{0} &\displaystyle{r < 4 R_{\rm S}}
        \\[6pt]
        \displaystyle{\rho_{\rm sp}^{7/3}(r)} & \displaystyle{4 R_{\rm S} \leq r \leq R_{\rm sp}}
        \\[10pt]
        \displaystyle{\rho_{\rm NFW}(r)} & \displaystyle{r \geq R_{\rm sp}}
    \end{cases} ~,
    \label{equ:dm_distro_adiabatic}
\end{equation}
where $\rho^{7/2}_{\rm sp}$ is the DM density with the spike parameter $\gamma_{\rm sp} = 7/2$.
There are multiple ways the above adiabatic accumulation of DM could be perturbed. For example, first, stars orbiting the central black hole can soften the spike via gravitational heating of the DM, and, if significant, can change the expected DM profile. This is reflected by the change in expected $\gamma_{\rm sp}$. For the MW with the profile given above,  $\gamma_{\rm sp} = 7/3\rightarrow \gamma_{\rm sp}=3/2$~\cite{Bertone:2005hw, Bertone:2005xv, Balaji:2023hmy, BetancourtKamenetskaia:2025ivl}. Secondly, if DM self-interacts, e.g. via annihilation, then the maximum of the spike would be reduced to $\rho_c$~\cite{Gondolo:1999ef, Balaji:2023hmy, Akita:2025dhg}. Here $\rho_{\rm c}= m_\chi/(\langle \sigma v \rangle t_{\rm BH})$ with $m_\chi$ the DM mass, $\langle \sigma v \rangle$ the thermally-averaged annihilation cross-section, and $t_{\rm BH}$ the age of the SMBH. Combined, the resulting DM distribution for the MW would be
\begin{equation}
    \rho_\chi^{\rm MW} (r) = 
    \begin{cases}
        \displaystyle{0} &\displaystyle{r < 4 R_{\rm S}^{\rm MW}}
        \\[6pt]
        \displaystyle{\frac{\rho_{\rm sp}^{3/2}(r) \rho_{\rm c}}{\rho_{\rm sp}^{3/2}(r) + \rho_{\rm c}}}&\displaystyle{4 R_{\rm S}^{\rm MW} \leq r \leq R_{\rm sp}^{\rm MW}}
        \\[10pt]
        \frac{\displaystyle{\rho_{\rm NFW}(r)} \rho_c}{\displaystyle{\rho_{\rm NFW}(r)} + \rho_c}&\displaystyle{r \geq R_{\rm sp}^{\rm MW}}
    \end{cases} ~,
    \label{equ:dm_distribution_mw}
\end{equation}
where $\rho^{3/2}_{\rm sp}$ stands for the DM density in the spike with $\gamma_{\rm sp} = 3/2$.
Note, while in the above equations we set the onset of the spike starting at $4R_\mathrm {S}$, the lower bound could be as small as $2R_{\mathrm{S}}$ after taking into account general-relativistic effects~\cite{Sadeghian:2013laa, Ferrer:2017xwm, Ferrer:2022kei}, but it will have a negligible impact on  our results.

\cref{fig:density} shows the DM density profiles for the MW used in this work. We set $M_{\rm BH} = 4 \times 10^{6}~M_{\odot}$, and $\gamma = 1$. Above one pc, the profile follows a standard NFW profile (green dashed), whereas below one pc, the spike takes over (solid black). The dotted line denotes the effects of self-annihiltion when setting $\langle \sigma v\rangle = 10^{-25}\;\mathrm{cm^3 / s}$, while the black dashed line denotes the density profile when including stellar heating~\cite{Balaji:2023hmy}.

In the following,  we discuss CR acceleration within MW-like galaxies ($M_\mathrm{BH}\sim 4\times 10^{6}M_\odot$~\cite{GRAVITY:2023avo}), and NGC 1068-like galaxies ($M_\mathrm{BH}\sim 8\times 10^{6}M_\odot$~\cite{Lodato:2002cv}). This puts them at the low end of expected masses for SMBHs in AGNs, where we expected ranges from $10^6 M_\odot- 10^{10} M_\odot$~\cite{Woo:2002un, Dalla_Bont__2025}. Since the spike density scales with the mass of the SMBH as $\sim M^{0.7}$, using low masses, as we do here, should be considered a conservative approach. The corresponding acceleration regions are shown by the pink vertical lines in Figure~\ref{fig:density}.


\section{DM Cosmic Ray Interactions}
\label{sec:dm_cr_interactions}

There are two relevant quantities needed for the discussion here. 
One is the momentum transfer during a CR and DM interaction event. 
The maximal momentum transfer can be written as~\cite{Ema:2018bih, Bringmann:2018cvk, Cappiello:2018hsu, Dent:2019krz, Ciscar-Monsalvatje:2024tvm}
\begin{equation}
    T_\mathrm{max} = \frac{T_\mathrm{CR}^2 + 2 m_\mathrm{CR}T_\mathrm{CR}}{T_\mathrm{CR} + (m_\mathrm{CR} + m_\chi)^2 / (2 m_\chi)}.
\end{equation}
Here we consider the energy transfer in the nucleus's rest frame, with $Q^2 = -t$, where $t$ is the Mandelstam variable. Setting $Q^2 = -(p_{\rm \chi} - p_{\chi}')^2 > 0$, we get $Q^2 = -2m_{\chi}^2 + 2 m_{\chi}E_{\chi}' = 2m_{\chi}T_{\chi}' < 2 m_\chi T_{\rm max}$. We now make the conservative requirement that $\sqrt{Q^2} \gtrsim 1$~GeV for iron to shatter, far above the binding energy of iron, $\epsilon_{b} \sim 9$~MeV, to ensure a thorough fragmentation. Therefore, requiring $m_\chi > 1 \text{ GeV}^2/2T_{\rm max} \simeq 1\text{ GeV}^2/2T_{\rm CR}$ implies that for $50$~EeV, we require $m_\chi \gtrsim 0.01$~eV.
For a more detailed discussion on momentum transfers and cross sections, see Refs.~\cite{Bardhan:2022bdg, Zhang:2025rqh}.

We also define the interaction length, $\lambda_{\chi, \mathrm{CR}}$ as
\begin{equation}
    \lambda_{\chi, \mathrm{CR}} = \frac{1}{n_\chi \sigma_{\chi\mathrm{CR}}},
\end{equation}
where $n_\chi$ is the number density of DM and $\sigma_{\chi\mathrm{CR}}$ the DM-Nucleus cross section.

Note that at the energies discussed here, we are beyond the regime of elastic scattering. 
For instance, for the scattering of a $T_\mathrm{iron} = 100$ TeV iron nucleus on a $m_\chi = 0.1$ MeV DM target, the average momentum transfer is $\langle{\Delta T}\rangle \sim 4.5$ GeV, demonstrating that we are safely within the regime of deep inelastic scattering.


\section{Iron Nuclei Fragmentation}
\label{sec:bounds}

Here we focus on the possibility of accelerating iron to ultra-high energies near the SMBH and/or the inner jet, assuming a background DM distribution with couplings to protons. See Refs.~\cite{Rieger:2019wgt, Rieger:2022qhs} for reviews of possible models. 
Specifically, we ask the question: \textit{Would a coupling to DM shatter the iron nucleus before it can accelerate to observed ultra-high energies?}

\subsection{Accelerating within an AGN}

Assuming that UHECRs near the GZK cut-off ($E_\mathrm{CR}\sim 50\;\mathrm{EeV}$) are all accelerated within AGNs, we can set a basic requirement on the interaction length for CR-DM interactions, $\lambda_{\mathrm{CR},\chi}$. The interaction length $\lambda_{\mathrm{CR},\chi}$ needs to be \textit{larger} than the acceleration distance $l_\mathrm{acc}$ of the CR. Otherwise, the CR (at these extreme energies assumed to be iron) would shatter and could not be accelerated to the necessary energies. Conservatively, we can set $l_\mathrm{acc}$ to the size of the AGN. Current observations set the size of AGNs to be $\lesssim 10^{-3}\;\mathrm{pc}$~\cite{2022MNRAS.511.3005J, 2022ApJ...929...19G}. Note here we define the size of the AGN to be the size of the accretion disk of the central SMBH.

Thus, assuming CRs are accelerated within the accretion disk of the SMBH, we can set a conservative bound by setting $\lambda_{\chi, \mathrm{CR}} \geq 10^{-3}\;\mathrm{pc}$. Using the previously defined DM-density profiles (at $r=10^{-3}\;\mathrm{pc}$), we can then set a generic bound on $\sigma_{\chi, \mathrm{CR}}$
\begin{equation}
    \sigma_{\chi, \mathrm{CR}} \leq \left\{ 
    \begin{array}{ll}
    1.2\times 10^{-24}\left(\frac{m_\chi}{\mathrm{GeV}}\right)\;\mathrm{cm^2}     &  \mathrm{for~NFW} \\[6pt]
    1.3\times 10^{-28}\left(\frac{m_\chi}{\mathrm{GeV}}\right)\;\mathrm{cm^2} & \mathrm{for~Spike}
    \end{array}
    \right. \, .
    \label{eq:est1}
\end{equation}
While these bounds are negligible for DM masses above 1 keV compared to current constraints~\cite{LZ:2025iaw, Nadler:2019zrb}, for lower masses they begin to \textit{dominate over} current direct detection constraints.

\subsection{Accelerating within its trapped region}

We can refine the estimate in \cref{eq:est1} by observing that if a particle escapes its acceleration region, it no longer accumulates energy. In other words, the Larmor radius $r_L = E / (Z e B)$ needs to be smaller than the characteristic size of the source, $L$. Here, $B$ is the magnetic field strength within the source, $e$ the unit electric charge, and $Z$ the atomic number of the CR nucleus. This is called the \textit{Hillas Criterion}~\cite{Hillas:1984ijl}. As discussed in Refs.~\cite{Hillas:1984ijl, Ptitsyna:2008zs, Kotera:2011cp, Rieger:2022qhs} this is a \textit{necessary} but not sufficient constraint on CR acceleration. Thus, we can treat it as a conservative scale. The criterion can be written as
\begin{equation}
    E_\mathrm{CR} < (10^{20}\;\mathrm{eV}) Z \left(\frac{B}{10\,\mu\mathrm{G}}\right)\left(\frac{L}{10\;\mathrm{kpc}}\right).
\end{equation}
Note that this criterion can be improved by introducing flow velocities within the object~\cite{Ptitsyna:2008zs, Kotera:2011cp, Bell:2013vxa, Rieger:2022qhs}. Here, we change the criterion to the following: \textit{The particle's interaction length, $\lambda_{\chi, \mathrm{CR}}$, needs to be larger than the Larmor radius, $r_L$}. 
This is equivalent to requiring that the CR does not interact before escaping the acceleration site. 
Following Ref.~\cite{Rieger:2022qhs},  we define $r_L$ as
\begin{equation}\label{eq:larmor_scaling}
    r_L = \left(\frac{1.08\; \mathrm{pc}}{Z}\right)\left(\frac{E}{1\; \mathrm{PeV}}\right)\left(\frac{1\; \mu G}{B}\right) \, .
\end{equation}
To be conservative, we impose our criterion using the smallest Larmor radius imaginable at the source, namely that obtained with the magnetic field strength at the horizon of the SMBH. Additionally, we set the acceleration to be maximally efficient, such as in magnetic reconnection~\cite{Fiorillo:2023dts, Karavola:2024uui}, meaning the nucleon can accelerate fully within one $r_L$. 

From observations, the magnetic field strength at the SMBH's horizon was estimated to be $B\sim 10^4\;\mathrm{G}$~\cite{2009A&A...507..171S, Baczko:2016opl, Piotrovich:2020ooz}.
Thus, the Larmor radius for iron nuclei at the GZK cut-off in an AGN is
\begin{equation}\label{eq:larmor}
    r_L \approx 2\times 10^{-7}\;\mathrm{pc} \approx 6\times 10^{11}\;\mathrm{cm}.
\end{equation}
Note that $r_L$ in \cref{eq:larmor} is far smaller than the containment criterion in \cref{eq:est1} that used $r = 10^{-3}\;\mathrm{pc}$. 
On the other hand, the Larmor radius does not define \textit{where} the acceleration of the CR happens. We set the acceleration site to  $2\times 10^{-5}\;\mathrm{pc}$, which is larger than the maximum of 100$R_S$ (for the MW) used in the modelling of AGNs~\cite{Ptitsyna:2015nta, Murase:2019vdl, Inoue:2019yfs, Katsoulakos:2020rkr, Kheirandish:2021wkm, Eichmann:2022lxh, Inoue:2022yak, Fiorillo:2024akm}. 
We then require that the interaction length of Iron-DM is larger than the Larmor radius.

Another reason why this should be seen as a conservative bound is that it assumes the acceleration efficiency to be maximal, $\eta = 1$. 
As discussed in Ref.~\cite{Aharonian:2002we}, $\eta$ within these objects can be as low as $\eta \geq 0.03$. Note, other authors (e.g., Ref.~\cite{2009JCAP...11..009L}) describe this efficiency in terms of the acceleration time, $t_\mathrm{acc}$. 
In general, it can be written as
\begin{equation}
    t_\mathrm{acc} = \mathcal{A}t_\mathrm{L},
\end{equation}
with $\mathcal{A}$ being a multiplier of the Larmor time $t_\mathrm{L}$. 
In general, $\mathcal{A}\gg 1$ and in the specific case $\mathcal{A} \approx 1$, this corresponds to a maximally efficient acceleration process as seen in magnetic reconnection~\cite{2009JCAP...11..009L, Fiorillo:2023dts, Karavola:2024uui} (the assumption we are using here). In this scenario $t_\mathrm{acc}\sim t_\mathrm{scattering}$, where $t_\mathrm{scattering}$ is the scattering time-scale. 
We now apply these criteria and the resulting bounds are
\begin{equation}
    \sigma_{\chi, \mathrm{CR}} \leq \left\{ 
    \begin{array}{ll}
    1\times 10^{-22}\left(\frac{m_\chi}{\mathrm{GeV}}\right)\;\mathrm{cm^2}     &  \mathrm{NFW,} \\ [6pt]
    7\times 10^{-29}\left(\frac{m_\chi}{\mathrm{GeV}}\right)\;\mathrm{cm^2} & \mathrm{Spike,}
    \end{array}
    \right.
\end{equation}
improving the previous estimate for the spike, but reducing it for an NFW profile.

These bounds can now be further improved by assuming all CR are accelerated within an NGC 1068-like galaxy, instead of a MW-like one. Plugging in the numbers from Ref.~\cite{Herrera:2023nww} (and using the assumed emission region $\leq 10^{-4}\;\mathrm{pc}$) we get a significant improvement of the bounds to
\begin{equation}
    \sigma_{\chi, \mathrm{CR}} \leq 3\times 10^{-31}\left(\frac{m_\chi}{\mathrm{GeV}}\right)\;\mathrm{cm^2}.
\end{equation}

\subsection{Can the Cosmic Ray escape the galaxy?}
 Ref.~\cite{Gondolo:1999ef} predicts that the DM density rapidly falls for $r <10 R_\mathrm{S}$ and vanishes completely for $r < 4 R_\mathrm{S}$. For Sagittarius A$^*$ with $M = 4\times 10^{6}M_\odot$ the Schwarzshild radius is $3.8\times 10^{-7}$ pc. Thus, CRs could be accelerated fully within a high-density shell of DM. This opens up a new question: \textit{Does DM induce a GZK-like bound for CRs?} For simplicity, we alter this question and ask: \textit{Does DM induce a high-energy cutoff for iron nuclei?}
Since CRs are accelerated within 100$R_S <2\times10^{-5}\;\mathrm{pc}$ for a MW-like galaxy~\cite{Ptitsyna:2015nta, Murase:2019vdl, Inoue:2019yfs, Katsoulakos:2020rkr, Kheirandish:2021wkm, Eichmann:2022lxh, Inoue:2022yak, Fiorillo:2024akm} ($<1\times10^{-4}\;\mathrm{pc}$ for NGC 1068-like galaxies), we can ask if an iron nucleus can pass through the DM halo without a single interaction. This results in an upper-bound on $\sigma_{\chi p}$ of

\begin{equation}\label{eq:escape_bound}
    \sigma_{\chi, \mathrm{CR}} \leq \left\{ 
    \begin{array}{ll}
    3\times 10^{-24}\left(\frac{m_\chi}{\mathrm{GeV}}\right)\;\mathrm{cm^2}     &  \mathrm{ Burkert,} \\[6pt]
    3\times 10^{-26}\left(\frac{m_\chi}{\mathrm{GeV}}\right)\;\mathrm{cm^2}     &  \mathrm{NFW,} \\[6pt]
    9\times 10^{-31}\left(\frac{m_\chi}{\mathrm{GeV}}\right)\;\mathrm{cm^2} & \mathrm{MW~Spike,} \\[6pt]
    1\times 10^{-33}\left(\frac{m_\chi}{\mathrm{GeV}}\right)\;\mathrm{cm^2} & \mathrm{NGC~Spike}.
    \end{array}
    \right.
\end{equation}
This can be understood by comparing these numbers to the expectations from the GZK cut-off. For the GZK limit the photon-CR cross-section $\sigma_{\gamma p} \sim 3\times10^{-28}\;\mathrm{cm^{2}}$ is the relevant quantity~\cite{Arkhipov:2006vy}. The CMB denstiy is approximately 411 photons/$\mathrm{cm}^{3}$. This implies an interaction length of $\sim 10$ Mpc. Here the high DM densities scale the resulting cross-section for DM to lower values. 
For instance, the average density in a DM spike for a MW-like galaxy between $2\times10^{-5}\;\mathrm{pc}$ and $1$ pc is $\sim 7\times 10^9\;\mathrm{GeV / cm^3}$.

Note that the bound set by escape is stronger than the bound due to acceleration, mainly due to our conservative assumptions on the Larmor radius and acceleration time. Had we used smaller magnetic field strengths, the constraint from the trapped region would be stronger. For example, setting $B = 100\;\mathrm{G}$, results in $r_L \sim 4\times 10^{-5}\;\mathrm{pc}$, similar in size to the assumed location of acceleration. This alone would yield a bound due to the acceleration criteria similar to the escape shown in~\cref{eq:escape_bound}. If the acceleration were not 100\% efficient, this would only further improve.

\subsection{A non-conservative estimate}

While in the previous sections, we used conservative estimates to derive bounds on DM-CR interactions, here we relax them: In NGC 1068-like galaxies, CRs can be accelerated as close as $\sim 10^{-5}$ pc from the central SMBH~\cite{Herrera:2023nww}. Additionally, as discussed previously, the efficiency of acceleration could be as low as $\eta = 0.03$~\cite{Aharonian:2002we} (see Ref.~\cite{Yang:2024bsf} for even lower efficiencies). With these numbers $r_L \approx 1.3\times 10^{-5}$ pc and $\rho_\chi^{\rm NGC} (r=10^{-5}\;\mathrm{pc}) \approx 5\times 10^{18}\;\mathrm{GeV / cm^3}$. In this scenario, the acceleration bound is the dominant contribution and the resulting constraint on the cross-section is $\sigma_{\chi, {\rm CR}} \leq 8\times 10^{-35}\;\left(\frac{m_\chi}{\mathrm{GeV}}\right)\;\mathrm{cm^2}$.


\subsection{Baryonic Feedback}

Here we provide a rough estimate of baryonic feedback caused by the $\chi$-$p$ coupling on the formation of the spike itself. 
Baryonic feedback is relevant when DM has scattered at least once during the lifetime of the galactic DM halo, $t_\mathrm{age}$. 
In that case, in order to avoid baryonic feedback, we can require that
\begin{equation}
    t_\mathrm{age}\times n_p\times v_p \times \sigma_{\chi p} \leq 1.
\end{equation}
For illustration, we can use the age of the MW halo, $t_\mathrm{age}\sim 10\;\mathrm{Gyr} \sim 3\times 10^{17}$ seconds. For the baryon (proton) number density, $n_p$, we consider the stellar mass density, within the central parsec $\rho_\star \sim 10^{6}$--$10^{7}\,M_\odot/\mathrm{pc}^{3}$
($\sim 4\times10^{7}$--$4\times10^{8}\,\mathrm{GeV/cm^{3}}$) \cite{Genzel:2010zy,Schodel:2007ru,Schodel:2008up}, the gas density in the inner parsec $n \sim 10$--$100/\mathrm{cm}^{3} \sim 10$--$100\,\mathrm{GeV/cm^{3}}$ \cite{Baganoff:2003jc}, and the central molecular zone $n \sim 10^{7}$--$10^{8}/\mathrm{cm}^{3} \sim 10^{7}$--$10^{8}\,\mathrm{GeV/cm^{3}}$~\cite{Mezger:1996,Genzel:2010zy}. For the velocity, we use $v_p \leq 0.1 c \simeq 3\times 10^9\;\mathrm{cm /s}$, which is above the observed velocity of stars close to Sagittarius A$^*$~\cite{Peissker:2020pse}. From this we obtain the approximate requirement that
\begin{equation}\label{eq:feedback_xsec}
    \sigma_{\chi p}\leq 10^{-35}\;\mathrm{cm^2}
\end{equation}
to avoid feedback.
The reasoning above assumes an isothermal core formation~\cite{BetancourtKamenetskaia:2025ivl} (see also Refs.~\cite{Spergel:1999mh, Kaplinghat:2015aga}).

\begin{figure}[t]
\centering
\includegraphics[width=\columnwidth]{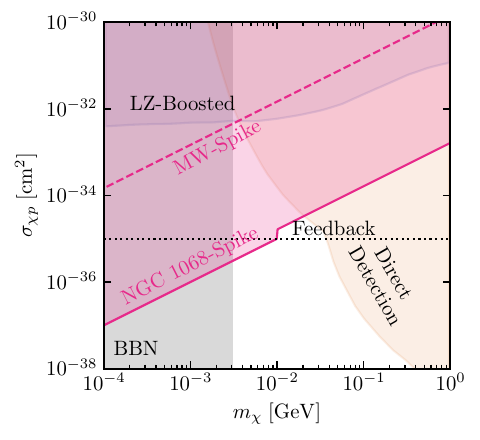}
\caption{The resulting bounds using the NFW and spiked profiles for the MW and assuming NGC 1068-like population. These bounds are constructed under the assumption of a GZK-like process. The experimental bounds are from LZ~\cite{LZ:2025iaw} CR-boosted DM search and the direct detection bounds adopted from Ref.~\cite{Billard:2021uyg} with data from CRESST~\cite{CRESST:2017ues, CRESST:2022dtl}, DAMIC~\cite{DAMIC:2020cut}, DarkSide~\cite{DarkSide:2018bpj, DarkSide:2018kuk}, and XENON1T\cite{XENON:2018voc, XENON:2019gfn}. The vertical shaded region is the BBN bound~\cite{Sabti:2019mhn, Krnjaic:2019dzc, Sabti:2021reh}.}
\label{fig:res_gzk}
\end{figure}

For larger cross sections (e.g., for most of the parameter space we target in the MW scenario), we assume the DM density profile follows the strong stellar heating scenario shown in \cref{fig:density} and calculated in Ref.~\cite{Balaji:2023hmy}. 
\cref{fig:res_gzk} shows the resulting bounds including baryonic feedback. For the MW scenario, this applies for all DM masses, while for the NGC 1068 case, masses $\geq 10\;\mathrm{MeV}$ are affected, which appears in our curves as a kink at about $m_\chi \sim 10$~MeV. We note that NGC 1068 is a significantly younger source, so the corresponding bound in \cref{eq:feedback_xsec} will weaken accordingly.

\cref{fig:res_gzk} shows the resulting bounds for spiked profiles, compared to current constraints from CR-boosted DM search~\cite{LZ:2025iaw}, direct detection searches~\cite{Billard:2021uyg} and BBN~\cite{Sabti:2019mhn, Krnjaic:2019dzc, Sabti:2021reh}.

\begin{figure}[t]
\centering
\includegraphics[width=\columnwidth]{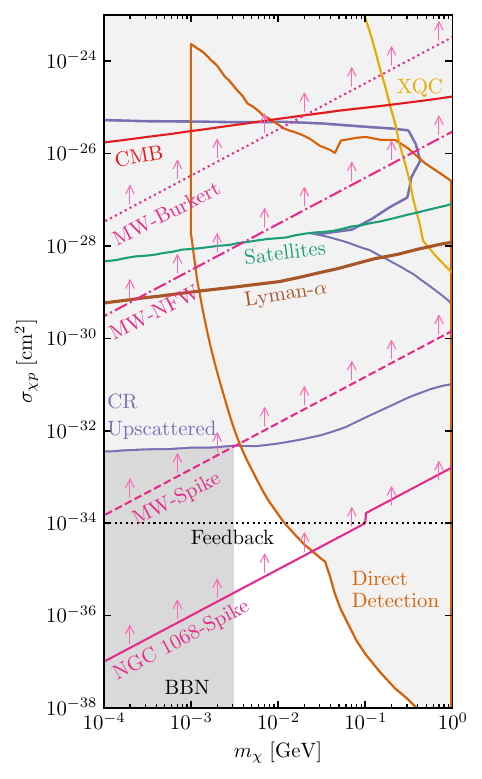}
\caption{
Summary of our bounds on DM-proton cross section (lines with upward arrows) compared to current constraints from XQC~\cite{Erickcek:2007jv}, CMB~\cite{Gluscevic:2017ywp}, upscattered DM (LZ-boosted~\cite{LZ:2025iaw} and PROSPECT~\cite{PROSPECT:2021awi}), EDELWEISS~\cite{EDELWEISS:2019vjv}, CRESST 2017 (surface)~\cite{CRESST:2017ues, Emken:2018run}, XENON1T~\cite{XENON:2017vdw, XENON:2018voc, XENON:2019gfn}, CRESST-III~\cite{CRESST:2019jnq, CRESST:2019axx}, DAMIC~\cite{DAMIC:2020cut}, and DarkSide~\cite{DarkSide:2018bpj, DarkSide:2018kuk},  BBN~\cite{Sabti:2019mhn, Krnjaic:2019dzc, Sabti:2021reh}, MW Sattelites~\cite{Nadler:2019zrb}, Lyman-$\alpha$~\cite{Rogers:2021byl}, and re-interpretations of existing data~\cite{Emken:2019tni}.}
\label{fig:res_gzk_tall}
\end{figure}

\cref{fig:res_gzk_tall} shows a summary of all relevant experimental constraints, compared to our bounds which rule out  new parameter space between in the DM mass region $m_\chi\in [3,30]$ MeV for the NGC 1068 case with a spiked profile.

\subsection{A comment on TXS 0506+056}

TXS 0506+056 was observed to be a source of neutrinos by Icecube~\cite{IceCube:2018cha}. Based on the neutrino and Gamma-Ray observations~\cite{2017ATel10791....1T, MAGIC:2018sak}, a few models predict that the main acceleration region of CR is $\mathcal{O}$(pc) away from the central SMBH~\cite{deGouveiaDalPino:2024txs, Kun:2018txs, Caproni:2022txs}, well outside of the possible DM spike. Specifically, in Ref.~\cite{Fiorillo:2025cgm} the authors show that a coronal origin (close to the SMBH) is disfavored by X-ray estimations. 

On the other hand, models exist where the production of neutrinos (thus the acceleration of CRs) happens between $4\times 10^{-4} - 0.3$ pc~\cite{Xue:2019txw, Xue:2020kuw, Yang:2024bsf}. Based on this, the spiked profile has been used to constrain DM-neutrino interactions in those scenarios~\cite{Cline:2022qld, Ferrer:2022kei, Zapata:2025huq}.

Setting the mass of the central SMBH to be $3\times 10^8\;M_\odot$~\cite{Padovani:2019xcv}, the DM density will be between $10^{10} - 10^{15}$ GeV/cm$^3$ for the distances of interest. Assuming all high-energy (iron) CRs are accelerated within TXS-like objects, this would lead to a bound of $\sigma_{\chi, {\rm CR}} \lesssim 10^{-33} \left(\frac{m_\chi}{\mathrm{GeV}}\right)$ cm$^2$, when considering that CR should be able to escape the spike without interacting. This is similar to the NGC bound and is mainly due to the similar column densities between the two when considering escaping CRs.

\section{Conclusion}
\label{sec:conclusion}

Considering the recent spike in interest in spiky dark matter density profiles around supermassive black holes, it is very timely to find new observables to test the existence of such DM spikes.
Conversely, it is also timely to ask what DM properties can be constrained in the presence of such spikes.
In this work, we have introduced a novel method to constrain the DM-nucleon cross section within the spikes under various assumptions. 
We noted that relatively weak interactions between DM and SM baryons has the potential to disrupt the acceleration of ultra-high-energy CRs in dense environments near SMBHs. 
This is the case even after accounting for baryonic feedback that can suppress DM spikes.

To set bounds on the values of these cross sections or to make statements on the existence of DM spikes, we relied on a couple of reasonable assumptions.
The first, and most crucial, is that CRs, particularly the heaviest elements like iron nuclei, are accelerated within DM spikes surrounding active galactic nuclei.
If that is the case, they would fragment before reaching observed ultra-high energies, for sufficiently large $\sigma_{\chi p}$.
By requiring that iron nuclei with energies $\sim 50$~EeV can be accelerated and escape without shattering, we derived conservative upper bounds on the DM–proton cross section. For a spiked DM profile in a MW-like galaxy, we obtained $\sigma_{\chi p} \lesssim 7 \times 10^{-29}(m_\chi/\mathrm{GeV})\mathrm{cm}^2$, improving to $\sigma_{\chi p} \lesssim 3 \times 10^{-31}(m_\chi/\mathrm{GeV})\mathrm{cm}^2$ when using NGC 1068-like galaxies.

We also examined whether DM could induce a GZK-like cut-off for iron nuclei within the AGN itself. 
From this, we set a bound on the DM–proton scattering cross section of $\sigma_{\chi p} \lesssim 9 \times 10^{-31}(m_\chi/\mathrm{GeV})\mathrm{cm}^2$ assuming a spiked profile for MW-like galaxies and $\sigma_{\chi p} \lesssim 1 \times 10^{-33}(m_\chi/\mathrm{GeV})\mathrm{cm}^2$ for NGC 1068-like galaxies.

These bounds probe DM interactions at momentum transfers $\gtrsim 10$~MeV, offering complementary and, in some cases, stronger constraints than those from direct detection or cosmological observations. 
In addition, given the large momentum transfers probed in this system, our bounds are also relevant to inelastic DM scenarios~\cite{Tucker-Smith:2001myb}, where the interactions of DM particles with nucleons are necessarily inelastic, requiring a minimum amount of momentum exchange to produce the excited DM state, $\chi p \to \chi^*p$.
These scenarios are much less constrained by direct detection experiments, though model-dependent limits from accelerators also apply~\cite{Izaguirre:2015zva,Berlin:2018jbm} (see also Ref.~\cite{Acevedo:2025rqu}).
Our results demonstrate that ultra-high-energy CR survival and multi-messenger signatures from AGNs can provide powerful and model-independent constraints on DM.
We emphasize that more robust and improved limits can be obtained by incorporating DM spikes in more detailed CR acceleration models and studying the energy dependence of the composition of CRs as observed at Earth.


\section*{Acknowledgments}

We are grateful for helpful discussions with Chris Cappiello, Dibya Chattopadhyay, Peter Cox, Damiano Fiorillo, Raj Gandhi, Avirup Ghosh, Jayden Newstead and Yago Porto. This work was supported by the Australian Research Council through Discovery Project DP220101727 plus the University of Melbourne’s Research Computing Services and the Petascale Campus Initiative.
This work was partially supported by the University of Iowa’s Year 2 P3 Strategic Initiatives Program through funding received for the project entitled ``High Impact Hiring Initiative (HIHI): A Program to Strategically Recruit and Retain Talented Faculty.'' The work of B.D. was partly supported
by the US Department of Energy under grant No. DE-SC0017987 and by a Humboldt
Fellowship from the Alexander von Humboldt Foundation. S.A.M. and B.D. wish to acknowledge the Center for Theoretical Underground Physics and Related Areas (CETUP*) and the Institute for Underground Science at SURF for hospitality and for providing a stimulating environment, where a part of this work was done.  B.D. also thanks the organizers of WHEPP 2025 at IIT, Hyderabad for local hospitality (and for arranging hospital visits) during the final stages of this work.  

\appendix
\counterwithin{figure}{section}


\section{Is there a spike?}\label{app:arguments}

The existence of a DM spike is not well-established in the literature; see for example, the arguments for and against it in  Refs.~\cite{Ullio:2001fb, Merritt:2002vj, Merritt:2003eu}. Specifically, one of the mechanisms for the destruction of DM spikes are (galactic) mergers. 
During such mergers, the spike can be tidally disrupted reducing the density and consequently $\gamma_\mathrm{sp}$.
On the other hand, Ref.~\cite{Bertone:2002je} argues that specifically for the MW, this destruction mechanism is unlikely, mainly due to the fact that the center of the MW is believed to have formed 12 billion years ago in a single merger, without subsequent ones.

Another option for the suppression of the spike is stellar heating and dynamical friction~\cite{Merritt:2003qk}. 
While the DM spike is partially argued to exist due to the steeply rising stellar density near the galactic center, these stars could in turn lower the density via kinetic heating and capture.
These processes would shift the expected $\gamma_\mathrm{sp}\sim 7/3$ to $3/2$. 
While this is above the value predicted by the NFW profile, it is below the adiabatic prediction.
Of note, is that observationally, there are claims that $\ \gamma\sim2.3$~\cite{chan2024first}, which would be closer to a purely adiabatic regime.
There are also suggestions, that the growth could happen instantaneously, not adiabatically. In such a scenario, while there would be a spike, it would be ``blunted" compared to the adiabatic regime~\cite{Ullio:2001fb}.
We note that even though stellar heating could suppress the spike~\cite{DelaTorreLuque:2024wfz}, even in the maximal heating scenario, the DM density is still well above the predictions of a pure NFW profile.

The main issue with the spiked profile is the Core-Cusp problem. 
The prediction of the spike builds on a cuspy (NFW) profile. 
Some measurements, most notably the recent Gaia data~\cite{Ou:2023adg, Lim:2023lss, Baumgart:2025dov} seem to suggest that the DM density at the center of MW is more cored, which would significantly reduce the DM spike. 
However, there are no measurements nor simulations below kpc-scale to distinguish a spiked profile from a cored one. Moreover, even though the MW may not host a spiked DM profile, AGNs with ongoing accretion and adiabatic growth of the SMBH are more likely to do so.

\bibliography{bibliography}


\end{document}